\documentclass[a4paper,12pt]{article}

\usepackage{ifpdf}

\newif\ifpdf
\ifx\pdfoutput\undefined
  \pdffalse
\else
  \pdfoutput=1
  \pdftrue
\fi

\RequirePackage{xspace} %
\RequirePackage{subfigure} %
\RequirePackage[centertags]{amsmath} %
\RequirePackage{amssymb}
\RequirePackage{wrapfig} %
\RequirePackage{calc} %
\RequirePackage{ifthen}
\RequirePackage{tabularx} %
\RequirePackage{flafter} %
\RequirePackage{fancyhdr} %

\ifpdf
  \RequirePackage[pdftex]{color}%
  \RequirePackage{colortbl}%
  \RequirePackage{array}%
  \RequirePackage[pdftex]{graphicx}

  \RequirePackage[ pdftex, plainpages = false, pdfpagelabels,
                 pdfpagelayout = useoutlines,
                 bookmarks,
                 breaklinks = true,
                 linktocpage,
                 pagebackref,                      
                 colorlinks = true,
                 linkcolor = blue,
                 urlcolor  = blue,
                 citecolor = blue,
                 anchorcolor = blue,
                 hyperindex = true,
                 hyperfigures
                 ]{hyperref}

\else
  \RequirePackage{color}
  \RequirePackage{colortbl}
   \RequirePackage{array}
  \RequirePackage[dvips]{graphicx}
  \RequirePackage{hyperref}
  \usepackage{rotating}
\fi

\usepackage{makeidx} 
\usepackage{setspace} 
\usepackage{rotating} 
\usepackage{ecltree}
\usepackage{epic}
\usepackage{supertabular}  
\usepackage{color}
\usepackage{exscale}
\usepackage{fontenc}
\usepackage{ifthen}
\usepackage{latexsym}
\usepackage{makeidx}
\usepackage{syntonly}
\usepackage{inputenc}
\usepackage{graphicx}
\usepackage{setspace}
\usepackage{caption2}
\usepackage[english]{babel}
\usepackage[square, comma,numbers,sort&compress]{natbib}
\usepackage{hypernat}
\usepackage{boxedminipage}
\usepackage{framed}
\usepackage{longtable}
\usepackage[all]{hypcap}    
\usepackage{algorithm2e}
\usepackage{algorithmic}
\usepackage{lscape}
\usepackage{pdflscape}

\newcommand{\note}[1]{\marginpar[left]{\singlespace \tiny #1}}
\newcommand{\pois}{Poiseuille}
\newcommand{\Hs}      {\hspace{-0.5cm}} %
\newcommand{\CIF}     {\centering \includegraphics[width=2.7in]} %
\renewcommand{\sectionmark}[1]%
      {\markright{\thesection\ #1}} 

\renewcommand{\note}[1]{}

\onehalfspace 

\begin{document}
\begin{center}
{\Large The flow of power law fluids in elastic networks and porous media}
\par\end{center}{\Large \par}

\begin{center}
Taha Sochi
\par\end{center}

\begin{center}
{\scriptsize University College London, Department of Physics \& Astronomy, Gower Street, London,
WC1E 6BT \\ Email: t.sochi@ucl.ac.uk.}
\par\end{center}

\begin{abstract}
\noindent The flow of power law fluids, which include shear thinning and shear thickening as well
as Newtonian as a special case, in networks of interconnected elastic tubes is investigated using a
residual based pore scale network modeling method with the employment of newly derived formulae.
Two relations describing the mechanical interaction between the local pressure and local cross
sectional area in distensible tubes of elastic nature are considered in the derivation of these
formulae. The model can be used to describe shear dependent flows of mainly viscous nature. The
behavior of the proposed model is vindicated by several tests in a number of special and limiting
cases where the results can be verified quantitatively or qualitatively. The model, which is the
first of its kind, incorporates more than one major non-linearity corresponding to the fluid
rheology and conduit mechanical properties, that is non-Newtonian effects and tube distensibility.
The formulation, implementation and performance indicate that the model enjoys certain advantages
over the existing models such as being exact within the restricting assumptions on which the model
is based, easy implementation, low computational costs, reliability and smooth convergence. The
proposed model can therefore be used as an alternative to the existing Newtonian distensible
models; moreover it stretches the capabilities of the existing modeling approaches to reach
non-Newtonian rheologies.

\vspace{0.3cm}

\noindent Keywords: fluid mechanics; power law fluid; elastic networks; elastic porous media.

\par\end{abstract}

\begin{center}

\par\end{center}

\section{Introduction} \label{Introduction}

The flow of non-Newtonian fluids in networks of interconnected distensible conduits occurs in many
fluid dynamic systems especially the biological ones in living organisms. The flow of blood; which
is a complex suspension with non-Newtonian attributes that include deformation rate dependency,
yield stress, thixotropy and viscoelasticity; is an obvious example \cite{BaskurtM2003, ChenLW2006,
BodnarSP2011, SochiNonNewtBlood2013}.

Although there are some previous investigations related to the non-Newtonian flow effects in small
distensible branching flow ensembles \cite{LiepschM1984, PerktoldRF1991, ChenL2006}, we are not
aware of an attempt to deal with this issue in a direct, general and large scale approach. The
obvious reason is the theoretical and computational difficulties associated with modeling such a
complex phenomenon which contains multiple non-linearities related to the rheology of the conduit
wall as well as the rheology of the fluid itself plus fluid-structure interaction. The existing and
commonly used Navier-Stokes models, whether the one-dimensional or the multi-dimensional ones, and
their derivatives fall short of incorporating such effects due to their restriction to the
Newtonian flow. Incorporating non-Newtonian effects in such models does not only require major
model development but it also requires considerable approximations that normally compromise the
results.

In this paper, we investigate the flow of shear-thinning and shear-thickening time-independent
non-Newtonian fluids of power law type in elastic networks using newly derived formulae
\cite{SochiElasticTubeNew2014} in conjunction with the residual based pore scale network modeling
\cite{SochiPoreScaleElastic2013}. The Newtonian case is obtained as a special case by setting the
flow behavior index of the power law fluid to unity. This inclusion is confirmed by the results
obtained from specially derived formulae for the Newtonian case.

Based on this approach, the pressure and volumetric flow rate solutions in such networks are
obtained by implementing the characteristic flow equations for the particular fluid and the
defining mechanical properties of the network tubes in a residual based pore scale network modeling
code as described in detail in \cite{SochiPoreScaleElastic2013} and outlined in section
\ref{Method}. In our formulation and implementation, we use two elastic models to characterize the
mechanical response of the tube wall that correlates the magnitude of the local pressure to the
magnitude of the local cross sectional area.

In modeling the network flow, we adopt the same assumptions on which the derivation of the
analytical tube equations \cite{SochiElasticTubeNew2014} used to characterize the flow is based,
that is the flow in each tube is presumed incompressible, laminar, time-independent, fully
developed at comparatively low Reynolds numbers with minimal entry and exit edge effects. Hence,
the flow is regarded as essentially viscous with minimal inertial contributions.

Furthermore, the walls of the tubes are assumed to be thin, homogeneous, isotropic, of constant
thickness over the whole tube length with linear distensibility and negligible compressibility to
ensure that the same characteristic mechanical response applies to all cross sections at all axial
positions and in all radial directions leading to an axi-symmetric expansion in the tube geometry.
We also rule out the possibility of a collapsed tube by imposing the condition that the
pressure-modified cross sectional area cannot be smaller than its reference state corresponding to
the reference pressure where the tube assumes its regular cylindrical shape under infinitesimal
stress.

Although the individual tubes in the network can in principle have different pressure-area
mechanical characteristics, for the sake of simplicity the results generated and reported in this
paper are obtained with networks in which all the tubes are assumed to have the same mechanical
characteristics. This is normally the case in the most common types of synthetic flow networks. It
is also realistic in most biological networks, such as blood circulation system, as long as the
modeled network does not span different categories of vessels belonging to different circulation
subsystems with different mechanical characteristics, e.g. arteries and veins. Anyway, the model
can be easily modified to accommodate multiple tube wall mechanical characteristics if such a
modification is required.

As done in the past by many network modelers \cite{SorbieCJ1989, OrenBA1997, Valvatnethesis2004,
Balhoffthesis2005, SochiThesis2007}, the network models can also be used as prototypes for porous
media. This can be very good approximation when the void space is characterized by rather regular
patterns with statistical distributions that can be well averaged and described by the distribution
of the tubes in the network.

\section{Method}\label{Method}

According to the residual based pore scale network modeling approach, the flow rate and pressure
fields in a fully connected network of tubes with certain mechanical properties that is filled with
a particular fluid and subjected to a pressure gradient can be obtained by imposing the presumed
boundary conditions on the inlet and outlet boundary nodes and enforcing a flow continuity
condition on each interior node of the network using a characteristic flow relation. This relation
normally correlates the volumetric flow rate to the boundary pressures in a single conduit and is
based on the particular rheological fluid model and the characterizing properties of the flow
conduit such as the flow of a power law fluid in a tube of elastic nature. Since in most cases
these problems are non-linear requiring the solution of a system of nonlinear simultaneous
equations, a non-linear iterative solution scheme, such as Newton-Raphson method, is then employed
by starting from an initial guess for the flow and pressure fields and iterating, while trying in
each iteration to minimize the residual flow at each interior node by imposing the continuity
condition. The final solution is then obtained when two consecutive iterations converge to the same
solution within a given error tolerance. This method is fully described in
\cite{SochiPoreScaleElastic2013} for the case of Newtonian flow in distensible networks. Further
technical details about the non-linear residual based solution scheme can be found in
\cite{SochiTechnical1D2013}, although this reference is related to a one-dimensional Navier-Stokes
finite element flow model rather than a pore scale network model.

Regarding the analytical relations correlating the pressure to the volumetric flow rate in a single
tube that to be used in the proposed pore scale model for the flow of power law fluids in elastic
networks, they are given by the following equations as derived previously in
\cite{SochiElasticTubeNew2014}

\begin{equation}\label{plQ1}
Q=\left(\frac{\gamma
n^{n}\left[\left(\frac{p_{in}}{\gamma}+A_{o}\right)^{3(n+1)/2}-\left(\frac{p_{ou}}{\gamma}+A_{o}\right)^{3(n+1)/2}\right]}{3k\pi^{(n+1)/2}\left(3n+1\right)^{n}\left(n+1\right)L}\right)^{1/n}
\end{equation}
and

\begin{equation}\label{plQ2}
Q=\left(\frac{\beta
n^{n}\left[\left(\frac{A_{o}}{\beta}p_{in}+\sqrt{A_{o}}\right)^{(3n+2)}-\left(\frac{A_{o}}{\beta}p_{ou}+\sqrt{A_{o}}\right)^{(3n+2)}\right]}{2k\pi^{(n+1)/2}\left(3n+1\right)^{n}\left(3n+2\right)A_{o}L}\right)^{1/n}
\end{equation}
The first of these equations is based on a mechanical elastic response for the tube given by

\begin{equation}\label{pA1}
p=\gamma\left(A-A_{o}\right)
\end{equation}
while the second is based on an elastic tube model given by

\begin{equation}\label{pA2}
p=\frac{\beta}{A_{o}}\left(\sqrt{A}-\sqrt{A_{o}}\right)
\end{equation}
In the last four equations, $Q$ is the volumetric flow rate, $\gamma$ and $\beta$ are the stiffness
proportionality factors, $A$ is the tube cross sectional area corresponding to the actual pressure
$p$, $A_{o}$ is the reference cross sectional area corresponding to the reference pressure which
for simplicity is assumed to be zero, $p_{in}$ and $p_{ou}$ are the pressure at the tube inlet and
outlet respectively, $L$ is the tube length, and $k$ and $n$ are the power law consistency factor
and flow behavior index respectively.

\section{Implementation and Validation}

The network flow model described in this paper was implemented in a pore scale network modeling
computer code which is based on a Newton-Raphson iterative solution scheme that employs several
numerical solvers (UMFPACK, LAPACK, SPARSE, and SUPERLU). An extensive series of tests were then
carried out to inspect and assess the model. In the following we summarize some of the results of
these tests in the context of model validation. As we have no experimental data to compare with,
the validation tests are mainly focused on observing qualitatively and quantitatively correct
trends related to special and limiting cases where the right outcome can be predicted in advance.
As most of the validation tests rely on pore scale network flow models that are based on different
underlying characteristic flow equations, we outline these equations in the following paragraph.

In reference \cite{SochiElasticTubeNew2014}, the following two Equations for the flow of Newtonian
fluids in elastic tubes, whose derivation is based on the same method and assumptions used in the
derivation of the power law formulae (i.e. Equations \ref{plQ1} and \ref{plQ2}), were derived

\begin{equation}\label{NewtQ1}
Q=\frac{\left(p_{in}+\gamma A_{o}\right)^{3}-\left(p_{ou}+\gamma
A_{o}\right)^{3}}{24\pi\mu\gamma^{2}L}
\end{equation}
and

\begin{equation}\label{NewtQ2}
Q=\frac{\beta}{40\pi\mu
A_{o}L}\left[\left(\frac{A_{o}}{\beta}p_{in}+\sqrt{A_{o}}\right)^{5}-\left(\frac{A_{o}}{\beta}p_{ou}+\sqrt{A_{o}}\right)^{5}\right]
\end{equation}
where $\mu$ is the fluid dynamic viscosity. These equations are based on Equations \ref{pA1} and
\ref{pA2} respectively. In our validation tests we used these two formulae that describe the
characteristic flow in the network tubes and whose derivation is completely independent from the
derivation of the power law formulae to test the behavior of the power law flow model in this
special case since the Newtonian fluids are a sub-category of the power law fluids corresponding to
$n=1$. We also used the following equation, which correlates the flow rate to the pressure drop for
the flow of power law fluids in rigid cylindrical tubes \cite{SochiThesis2007,
SochiVariational2013}, to characterize the flow in the network tubes in some of these tests

\begin{equation}\label{plQ}
Q=\frac{\pi n}{3n+1}\sqrt[n]{\frac{1}{2k}\frac{dp}{dx}}\left(\sqrt{\frac{A}{\pi}}\right)^{3+1/n}
\end{equation}
where $x$ is the tube axial coordinate. Finally, the well known \pois\ model was also used in one
of our validation tests.

The following points outline the correct trends that have been observed and hence been used to
validate the model where all these trends are verified for both $p$-$A$ elastic tube models as
given by Equations \ref{pA1} and \ref{pA2}

\begin{figure} [!h]
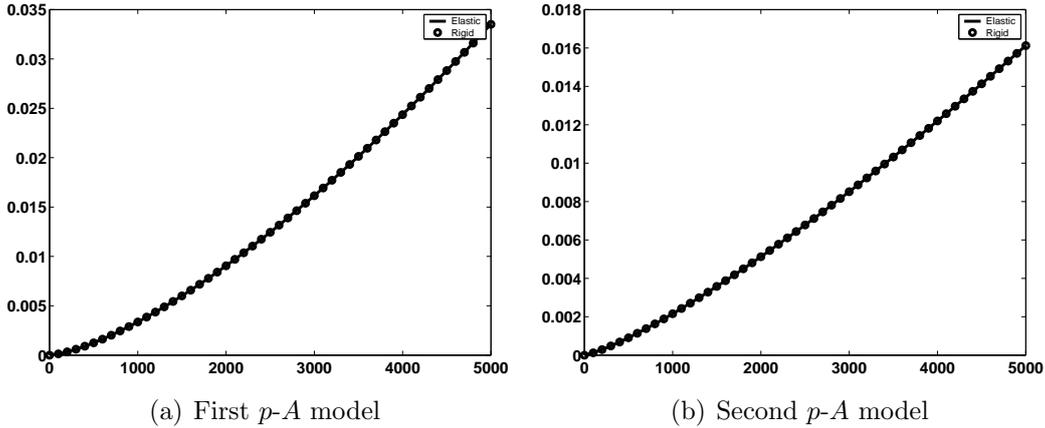

\centering %
\subfigure[First $p$-$A$ model]%
{\begin{minipage}[b]{0.5\textwidth} \CIF {g/PlotPlePlrComparisonPA1}
\end{minipage}}
\Hs %
\subfigure[Second $p$-$A$ model]%
{\begin{minipage}[b]{0.5\textwidth} \CIF {g/PlotPlePlrComparisonPA2}
\end{minipage}}
\caption{Convergence of solutions of power law pore scale elastic network model with high tube wall
stiffness (using Equations \ref{plQ1} and \ref{plQ2} for their single tube characteristic flow) to
their equivalents of rigid model (using Equation \ref{plQ} for single tube characteristic flow)
(a) for the first $p$-$A$ elastic model (Equation \ref{pA1}) with $\gamma=10^{12}$~Pa.m$^{-2}$
using a power law fluid with $n=0.7$ and $k=0.025$~Pa.s$^n$; and
(b) for the second $p$-$A$ elastic model (Equation \ref{pA2}) with $\beta=10^6$~Pa.m using a power
law fluid with $n=0.8$ and $k=0.015$~Pa.s$^n$.
In both cases a fractal network consisting of 511 tubes in 9 generations with an area-preserving
branching exponent of 2 \cite{SochiBranchFlow2013, SochiPoreScaleElastic2013} and an inlet main
branch with $L=0.08$~m and $R_o=0.016$~m was used. The vertical axis in the sub-figures represents
the net inflow/outflow, $Q$, in m$^3$.s$^{-1}$ while the horizontal axis represents the inlet
boundary pressure, $p_{in}$, in Pa. The outlet pressure in both cases is zero.
\label{PlotPlePlrComparison}}
\end{figure}

\begin{figure} [!h]
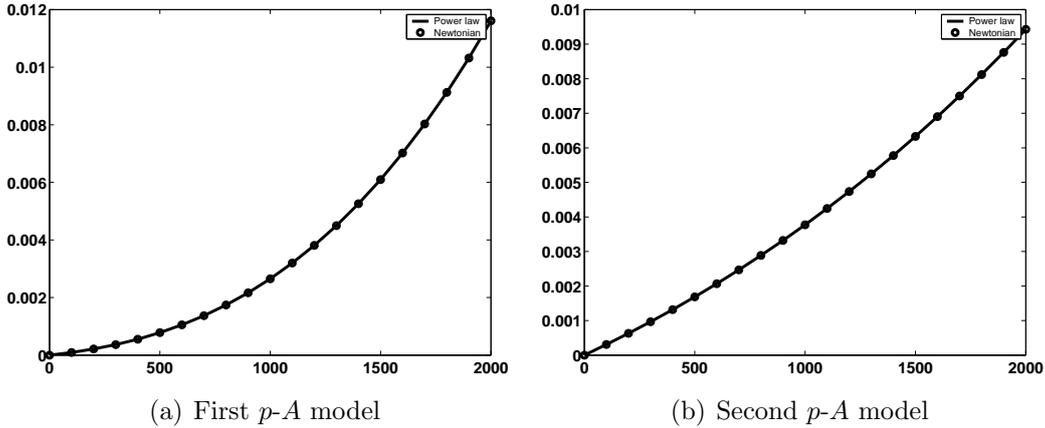

\centering %
\subfigure[First $p$-$A$ model]%
{\begin{minipage}[b]{0.5\textwidth} \CIF {g/PLotPleNewtComparisonPA1}
\end{minipage}}
\Hs %
\subfigure[Second $p$-$A$ model]%
{\begin{minipage}[b]{0.5\textwidth} \CIF {g/PLotPleNewtComparisonPA2}
\end{minipage}}
%
\caption{Convergence of solutions of power law elastic network model (using Equations \ref{plQ1}
and \ref{plQ2} for their single tube characteristic flow) to their equivalents of Newtonian elastic
model (using Equations \ref{NewtQ1} and \ref{NewtQ2} for their single tube characteristic flow)
when $n=1$
(a) for the first $p$-$A$ elastic model (Equation \ref{pA1}) with $\gamma=10^{7}$~Pa.m$^{-2}$ using
a fluid model with $k=\mu=0.03$~Pa.s; and
(b) for the second $p$-$A$ elastic model (Equation \ref{pA2}) with $\beta=80$~Pa.m using a fluid
model with $k=\mu=0.008$~Pa.s.
In both cases an orthorhombic network consisting of 946 tubes with an average $R_o$ of 0.0042~m and
a standard deviation of 0.0015~m \cite{SochiPoreScaleElastic2013} was used. The axes and outlet
pressure are as in Figure \ref{PlotPlePlrComparison}. \label{PLotPleNewtComparison}}
\end{figure}

\begin{figure} [!h]
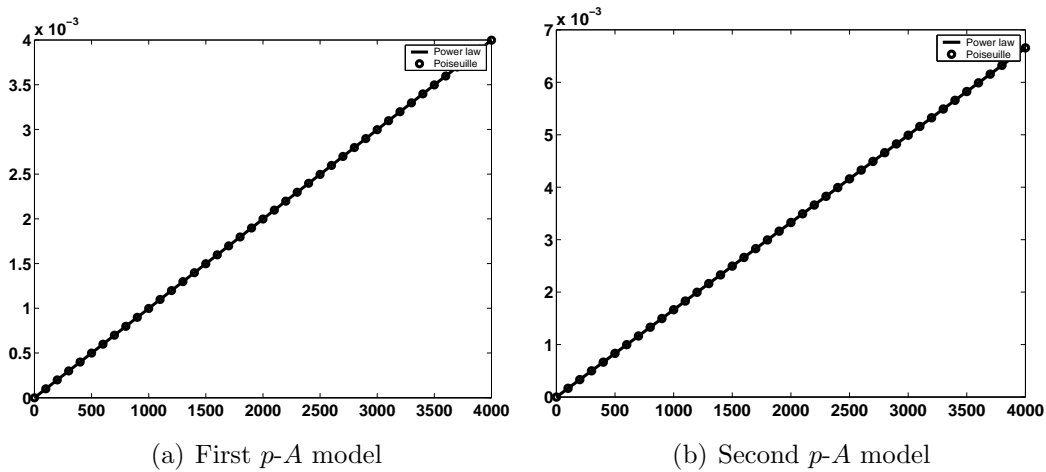

\centering %
\subfigure[First $p$-$A$ model]%
{\begin{minipage}[b]{0.5\textwidth} \CIF {g/PLotPlePoisComparisonPA1}
\end{minipage}}
\Hs %
\subfigure[Second $p$-$A$ model]%
{\begin{minipage}[b]{0.5\textwidth} \CIF {g/PLotPlePoisComparisonPA2}
\end{minipage}}
\vspace{-0.1cm} %
\caption{Convergence of solutions of power law pore scale elastic network model with high tube wall
stiffness and $n=1$ (using Equations \ref{plQ1} and \ref{plQ2} for their single tube characteristic
flow) to their equivalents of rigid \pois\ model
(a) for the first $p$-$A$ elastic model (Equation \ref{pA1}) with $\gamma=10^{11}$~Pa.m$^{-2}$
using a fluid model with $k=\mu=0.005$~Pa.s; and
(b) for the second $p$-$A$ elastic model (Equation \ref{pA2}) with $\beta=10^6$~Pa.m using a fluid
model with $k=\mu=0.003$~Pa.s.
In both cases a fractal network consisting of 255 tubes in 8 generations with a branching exponent
of 2.5 \cite{SochiBranchFlow2013, SochiPoreScaleElastic2013} and an inlet main branch with
$L=0.07$~m and $R_o=0.0105$~m was used. The axes and outlet pressure are as in Figure
\ref{PlotPlePlrComparison}. \label{PLotPlePoisComparison}}
\end{figure}

\begin{figure} [!h]
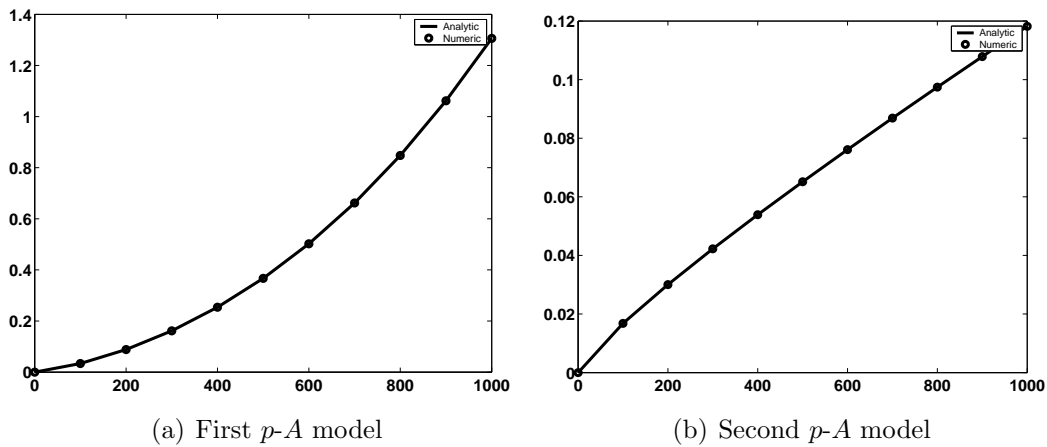

\centering %
\subfigure[First $p$-$A$ model]%
{\begin{minipage}[b]{0.5\textwidth} \CIF {g/PlotTubeNetworkComparisonPA1}
\end{minipage}}
\Hs %
\subfigure[Second $p$-$A$ model]%
{\begin{minipage}[b]{0.5\textwidth} \CIF {g/PlotTubeNetworkComparisonPA2}
\end{minipage}}
\caption{Comparison between the analytical solutions, as given by Equations \ref{plQ1} and
\ref{plQ2}, for a single elastic tube with $L=0.6$~m and $R_o=0.08$~m, and the numerical network
model solutions for the same tube obtained by discretizing the tube in the axial direction into 101
rings and treating the discretized ensemble as a network of serially connected tubes
(a) using the first $p$-$A$ elastic model (Equation \ref{pA1}) with $\gamma=50000$~Pa.m$^{-2}$,
$n=0.8$, and $k=0.2$~Pa.s$^n$; and
(b) using the second $p$-$A$ elastic model (Equation \ref{pA2}) with $\beta=7500$~Pa.m, $n=1.2$,
and $k=0.08$~Pa.s$^n$.
The axes and outlet pressure are as in Figure \ref{PlotPlePlrComparison}.
\label{PlotTubeNetworkComparison}}
\end{figure}

\begin{enumerate}

\item
The convergence of the solutions of the power law elastic network model (characterized by Equations
\ref{plQ1} and \ref{plQ2}) to the solutions of the power law rigid network model (characterized by
Equation \ref{plQ}) by increasing the tube wall stiffness. These trends are demonstrated in Figure
\ref{PlotPlePlrComparison} using a computer generated fractal network. Full description of this
type of networks is provided in \cite{SochiPoreScaleElastic2013}.

\item
The convergence of the solutions of the power law elastic network model (characterized by Equations
\ref{plQ1} and \ref{plQ2}) to the solutions of the Newtonian elastic model (characterized by
Equations \ref{NewtQ1} and \ref{NewtQ2} respectively) when $n=1$. These trends are demonstrated in
Figure \ref{PLotPleNewtComparison} using a computer generated orthorhombic network. Full
description of this type of networks is provided in \cite{SochiPoreScaleElastic2013}.

\item
As a consequence of the previous two points, it is expected that the solutions of the power law
elastic network model would converge to the solutions of the Newtonian rigid network model (\pois)
when $n=1$ and the stiffness of the tubes wall is high. This trend is also verified as demonstrated
in Figure \ref{PLotPlePoisComparison} using a computer generated fractal network.

\item
Another validation test is the agreement between the solutions of a discretized tube, as obtained
numerically from the residual based pore scale network modeling approach, and the analytical
solutions for a single tube, as given by Equations \ref{plQ1} and \ref{plQ2}. This is verified and
demonstrated in Figure \ref{PlotTubeNetworkComparison}. A similar test corresponding to the
Newtonian case has confirmed the agreement between the numerical solutions of the Newtonian elastic
network model and the single tube analytical expressions as given by Equations \ref{NewtQ1} and
\ref{NewtQ2}.

\end{enumerate}


\section{Assessment and Comparison}

It is difficult to make a fair comparison between two mathematical or computational models that are
based on different theoretical and numerical frameworks. This applies to any comparison between our
proposed pore scale network model and any one of the existing models such as the one-dimensional
Navier-Stokes flow model which is usually associated with a finite element numerical implementation
\cite{SochiTechnical1D2013}. In reference \cite{SochiElasticTubeNew2014}, we outlined some issues
related to such a comparison with regard to the flow in single tubes. Most of that discussion also
applies to the network flow models that are based on these two distinctive formulations.

However, there are some obvious advantages in using the pore scale network modeling approach over
the one-dimensional Navier-Stokes finite element model. Some of these advantages are based on our
personal experience and hence may be related to our computational frameworks for these models and
their particular implementations. The outcome may be different for other implementations although
we believe that the assessment is valid in general.

A major advantage of the proposed pore scale network modeling approach is the incorporation of
non-Newtonian effects, whereas the one-dimensional Navier-Stokes model is limited to the Newtonian
flow. Also, since the pore scale modeling approach is based on analytically derived characteristic
flow relations, it produces exact solutions within its defining assumptions considering that the
allowed error tolerance in the residual based iterative scheme can be minimized to the limit of
machine precision and hence can be forced to vanish for all practical purposes. Computationally, we
observe that the proposed pore scale modeling approach is superior in terms of robustness,
reliability, ease of implementation, low computational overhead, and better rate and speed of
convergence. For example, we experienced instabilities and convergence difficulties
\cite{SochiPois1DComp2013} when using the first $p$-$A$ relation with the one-dimensional
Navier-Stokes finite element flow model but not with the power law pore scale network model even
though the opposite should be expected since the latter employs a non-linear non-Newtonian rheology
while the former employs a linear Newtonian rheology. Similar assessment should also apply in
general to comparisons between the proposed pore scale network model and other existing models.

\section{Conclusions} \label{Conclusions}

The flow of non-Newtonian fluids in distensible networks and porous media is investigated where a
residual based pore scale network modeling method employing a recently derived formulae for the
flow of power law fluids in elastic cylindrical tubes is proposed. This modeling method was
implemented in a computer code and initial results based on extensive tests were obtained. The
results show a number of sensible trends which include convergence to special and limiting cases
where the outcome can be predicted from previously validated alternative models.

The investigation also included assessment of the proposed model and general comparison to the
existing ones notably the finite element based one-dimensional Navier-Stokes model. Several
advantages of the proposed model have been asserted such as producing exact solutions within its
underlying assumptions and allowed numerical errors, reliability, ease of implementation, low
computational cost, and fast and smooth convergence.

Although there are previous attempts where certain non-Newtonian effects were considered in the
flow through small scale distensible branching flow ensembles, the main novelty of the proposed
method is that it considers non-Newtonian rheology directly and on a large network scale employing
analytically derived underlying flow equations.

\vspace{1cm}
\phantomsection \addcontentsline{toc}{section}{Nomenclature} %
{\noindent \LARGE \bf Nomenclature} \vspace{0.5cm}

\begin{supertabular}{ll}
%
$\beta$                 &   stiffness factor in one of the pressure-area elastic models \\
$\gamma$                &   stiffness factor in another pressure-area elastic model \\
$\mu$                   &   fluid dynamic viscosity \\
\\
$A$                     &   tube cross sectional area \\
$A_o$                   &   tube reference cross sectional area at reference pressure \\
$k$                     &   consistency factor of power law fluid \\
$L$                     &   length of tube \\
$n$                     &   flow behavior index of power law fluid \\
$p$                     &   axial pressure \\
$p_{in}$                &   inlet pressure \\
$p_{ou}$                &   outlet pressure \\
$Q$                     &   volumetric flow rate \\
$R_o$                   &   radius of tube corresponding to tube reference area \\
$x$                     &   tube axial coordinate \\
\end{supertabular}

\phantomsection \addcontentsline{toc}{section}{References} %
\bibliographystyle{unsrt}

\begin{thebibliography}{10}

\bibitem{BaskurtM2003}
O.K. Baskurt;~H.J. Meiselman.
\newblock {Blood Rheology and Hemodynamics}.
\newblock {\em Seminars in Thrombosis and Hemostasis}, 29(5):435--450, 2003.

\bibitem{ChenLW2006}
J.~Chen; X-Y. Lu;~W. Wang.
\newblock {Non-Newtonian effects of blood flow on hemodynamics in distal
  vascular graft anastomoses}.
\newblock {\em Journal of Biomechanics}, 39:1983--1995, 2006.

\bibitem{BodnarSP2011}
T.~Bodn\'{a}r; A. Sequeira;~M. Prosi.
\newblock {On the shear-thinning and viscoelastic effects of blood flow under
  various flow rates}.
\newblock {\em Applied Mathematics and Computation}, 217(11):5055--5067, 2011.

\bibitem{SochiNonNewtBlood2013}
T.~Sochi.
\newblock {Non-Newtonian Rheology in Blood Circulation}.
\newblock {\em Submitted}, 2013.
\newblock arXiv:1306.2067.

\bibitem{LiepschM1984}
D.~Liepsch;~S. Moravec.
\newblock {Pulsatile flow of non-Newtonian fluid in distensible models of human
  arteries}.
\newblock {\em Biorheology}, 21(4):571--586, 1984.

\bibitem{PerktoldRF1991}
K.~Perktold; M. Resch;~H. Florian.
\newblock {Pulsatile Non-Newtonian Flow Characteristics in a Three-Dimensional
  Human Carotid Bifurcation Model}.
\newblock {\em Journal of Biomechanical Engineering}, 113(4):464--475, 1991.

\bibitem{ChenL2006}
J.~Chen; X-Y. Lu.
\newblock {Numerical investigation of the non-Newtonian pulsatile blood flow in
  a bifurcation model with a non-planar branch}.
\newblock {\em Journal of Biomechanics}, 39(5):818--832, 2006.

\bibitem{SochiElasticTubeNew2014}
T.~Sochi.
\newblock {The Flow of Newtonian and power law fluids in elastic tubes}.
\newblock {\em Submitted}, 2014.
\newblock arXiv:1405.4115.

\bibitem{SochiPoreScaleElastic2013}
T.~Sochi.
\newblock {Pore-Scale Modeling of Navier-Stokes Flow in Distensible Networks
  and Porous Media}.
\newblock {\em Computer Modeling in Engineering \& Sciences (Accepted)}, 2014.

\bibitem{SorbieCJ1989}
K.S. Sorbie; P.J. Clifford;~E.R.W. Jones.
\newblock {The Rheology of Pseudoplastic Fluids in Porous Media Using Network
  Modeling}.
\newblock {\em Journal of Colloid and Interface Science}, 130(2):508--534,
  1989.

\bibitem{OrenBA1997}
P.E. {\O}ren; S. Bakke;~O.J. Amtzen.
\newblock {Extending Predictive Capabilities to Network Models}.
\newblock {\em SPE Annual Technical Conference and Exhibition, San Antonio,
  Texas, SPE 38880}, 1997.

\bibitem{Valvatnethesis2004}
P.H. Valvatne.
\newblock {\em {Predictive pore-scale modelling of multiphase flow}}.
\newblock PhD thesis, Imperial College London, 2004.

\bibitem{Balhoffthesis2005}
M.T. Balhoff.
\newblock {\em {Modeling the flow of non-Newtonian fluids in packed beds at the
  pore scale}}.
\newblock PhD thesis, Louisiana State University, 2005.

\bibitem{SochiThesis2007}
T.~Sochi.
\newblock {\em {Pore-Scale Modeling of Non-Newtonian Flow in Porous Media}}.
\newblock PhD thesis, Imperial College London, 2007.

\bibitem{SochiTechnical1D2013}
T.~Sochi.
\newblock {One-Dimensional Navier-Stokes Finite Element Flow Model}.
\newblock {\em Technical Report}, 2013.
\newblock arXiv:1304.2320.

\bibitem{SochiVariational2013}
T.~Sochi.
\newblock {Using the Euler-Lagrange variational principle to obtain flow
  relations for generalized Newtonian fluids}.
\newblock {\em Rheologica Acta}, 53(1):15--22, 2014.

\bibitem{SochiBranchFlow2013}
T.~Sochi.
\newblock {Fluid Flow at Branching Junctions}.
\newblock {\em Submitted}, 2013.
\newblock arXiv:1309.0227.

\bibitem{SochiPois1DComp2013}
T.~Sochi.
\newblock {Comparing Poiseuille with 1D Navier-Stokes Flow in Rigid and
  Distensible Tubes and Networks}.
\newblock {\em Submitted}, 2013.
\newblock arXiv:1305.2546.

\end{thebibliography}

\end{document}